\documentclass[11pt]{article}

\usepackage[T1]{fontenc}
\usepackage{jheppub}
\usepackage{graphicx} 
\graphicspath{{Figures/}} 
\usepackage{hyperref}
\usepackage{mathtools}
\usepackage{amsmath}

\newcommand{\<}{\langle}
\renewcommand{\>}{\rangle}
\newcommand{\dd}{\mathrm{d}}
\newcommand{\Rbb}{\mathbb{R}}
\newcommand{\Pbb}{\mathbb{P}}
\newcommand{\Acal}{\mathcal{A}}
\newcommand{\Bcal}{\mathcal{B}}
\newcommand{\Ccal}{\mathcal{C}}
\newcommand{\Dcal}{\mathcal{D}}
\newcommand{\Ical}{\mathcal{I}}
\newcommand{\Ncal}{\mathcal{N}}
\newcommand{\Ocal}{\mathcal{O}}
\newcommand{\mat}[1]{\begin{pmatrix} #1 \end{pmatrix}}
\DeclareMathOperator*{\Res}{Res}

\title{An All-Loop Amplituhedron in Two Dimensions}

\author{Jonah Stalknecht}\emailAdd{jonah.stalknecht@matfyz.cuni.cz}

\affiliation{Institute for Particle and Nuclear Physics, \\ Charles University, \\ Prague, Czech Republic}

\abstract{
	We define and study a positive geometry $\Delta^{(L)}$ which serves as a natural generalization of loop amplituhedra to two-dimensional Minkowski space $\Rbb^{1,1}$. The geometry is formulated in the framework of lightcone geometries in dual momentum space, and can equivalently be obtained as a specific boundary of the $L$-loop amplituhedron for $\mathcal{N}=4$ super Yang--Mills. The simplicity of the two-dimensional setting allows us to calculate the canonical form of $\Delta^{(L)}$ at any loop order, which is shown to correspond to massless banana graphs. We integrate the canonical form at all loop orders in dimensional regularization, and find that the full IR divergence structure at $L$-loops is captured by the $L$\textsuperscript{th} power of the one-loop result, a phenomenon analogous to IR exponentiation. Furthermore, these integrated functions can be resummed into a closed-form non-perturbative result given by a Fox--Wright function. In the limit where $L\to\infty$, the geometry gives rise to a path integral over worldlines, suggesting the emergence of a dual description at strong coupling. This construction provides a simple and tractable setting in which to explore the geometry of loop amplitudes, and offers a controlled toy model for investigating loop amplituhedra beyond their standard scope.}

\begin{document}
	
\maketitle
%%%%%%%%%%%%%%%%%%%%%%%%%%%%%%%%%%%%%%%%
%%%%%%%%%%%%%%%%%%%%%%%%%%%%%%%%%%%%%%%%
\pagebreak

%%%%%%%%%%%%%%%%%%%%%%%%%%%%%%%%%%%%%%%%
\section{Introduction}

Loop amplituhedra provide a geometric framework to understand physics of loop integrands. The most important examples include the \emph{amplituhedron} \cite{Arkani-Hamed:2013jha} and the \emph{ABJM amplituhedron} \cite{He:2023rou}, which capture the planar integrands for scattering amplitudes in $\Ncal=4$ SYM and ABJM theory, respectively. Recently, it has been shown that these amplituhedra can be equivalently formulated as \emph{lightcone geometries} \cite{Lukowski:2023nnf, Ferro:2023qdp}: compact geometries in the Minkowski space of dual momentum variables, with boundaries given by lightcones. These lightcone geometries do not rely on the traditional momentum twistor variables, and are therefore more amenable to generalization to other dimensions or to massive particles. For the reader's convenience, we provide a review of positive geometries and lightcone geometries in appendix \ref{sec:appendix}.

In this paper, we will use the framework of lightcone geometries to define a new loop amplituhedron in two-dimensional Minkowski space $\Rbb^{1,1}$. The structure of lightcones simplifies drastically in two dimensions, which allows us to study the structure and canonical forms of these geometries at any loop order. We shall denote the $L$-loop lightcone geometry as $\Delta^{(L)}$. This geometry occupies a useful middle ground between physically interesting and mathematically simple. It captures many non-trivial physical properties of perturbative scattering amplitudes, many of which have direct analogues in $\Ncal=4$ SYM, while remaining simple enough that many all-loop and non-perturbative statements can be calculated exactly. Together, these features make it a useful toy model for investigating how the general amplituhedron framework encodes the physics of scattering amplitudes in regimes currently not accessible by other amplituhedra. We will investigate the geometry, canonical forms, and the integrated amplitudes described by $\Delta^{(L)}$. We note that the definition of the lightcone geometry $\Delta^{(L)}$ first appeared in the author's thesis \cite{Stalknecht:2024bdg}, though its properties and physical relevance were not studied there. 

\paragraph{Outline and Summary of Results.}
We will start with a brief review of kinematics in 1+1 dimensions and establish our conventions in section \ref{sec:kin}. We will define the lightcone geometry $\Delta^{(L)}$ in section \ref{sec:def}. Much like its three- and four-dimensional counterparts, we will see that $\Delta^{(L)}$ captures dual conformal invariant (DCI) $L$-loop integrands. Using lightcone coordinates, we calculate the canonical form of $\Delta^{(L)}$ at any loop order, and we find that it describes 2D massless \emph{banana graphs}. Furthermore, the geometry is naturally triangulated by $L!$ ``loop ordered'' regions, which only touch along codimension-2 boundaries. Each of these loop ordered regions is geometrically the Cartesian product of two $L$-simplices. The loop ordered regions only differ in permutations of the loop momenta, and as such it is sufficient to keep any one of these regions for the purpose of integration. 

In section \ref{sec:boundary}, we will see that $\Delta^{(L)}$ arises as a certain boundary of the 4-point $L$-loop amplituhedron for $\Ncal=4$ SYM. The boundaries which are shared by the loop-ordered regions are identified with the \emph{internal boundaries} of the amplituhedron discussed in \cite{Dian:2022tpf}. This further establishes $\Delta^{(L)}$ as a natural object to consider in the study of amplituhedra.

Although the amplituhedron framework usually stops at the integrand level, most of the physical interest is in the integrated amplitudes. In section \ref{sec:integrating} we will show that these massless banana graphs can be integrated exactly in dimensional regularization. We shall denote the result of integration as $A^{(L)}$. Similar to $\Ncal=4$ SYM, the DCI of the integrands is broken after integration. Although UV finite, $A^{(L)}$ has an IR divergence of order $1/\epsilon^L$. These are the direct analogue of ``soft-collinear'' singularities in $\Ncal=4$ SYM. We find that the full IR structure of $A^{(L)}$ is captured by the $L$\textsuperscript{th} power of $A^{(1)}$. Thus, this class of DCI integrands in 2D exhibit a type of \emph{IR exponentiation}. However, the non-perturbative sum does not exponentiate in the same way, and we instead find that the $L$-loop functions can be resummed into a \emph{Fox--Wright function} of the form $\phantom{}_2\Psi_1$, which is a generalization of a hypergeometric function. This resummation converges for $\epsilon<0$. To leading order, we find that this non-perturbative result is captured by a double pole structure in $A^{(1)}$:
\begin{align*}
    \sum_{L=0}^\infty g^L A^{(L)} \simeq \frac{1}{(1-\frac{g}{2}A^{(1)})^2}.
\end{align*}

In section \ref{sec:large-L} we will study the limit as $L$ goes to infinity, where the geometry $\Delta^{(L)}$ turns into the infinite-dimensional space of worldlines between two points in $\Rbb^{1,1}$. Upon considering the integral of $\Omega(\Delta^{(L)})$ in this limit, we will see the emergence of a path integral. This suggests the existence of a dual theory at strong coupling. 

We will round off with a conclusion and outlook in section \ref{sec:conclusions}.

%%%%%%%%%%%%%%%%%%%%%%%%%%%%%%%%%%%%%%%%
\section{Kinematics in Two Dimensions}\label{sec:kin}

Kinematics are highly constrained in two dimensions. It is familiar that any time-like vector in $\Rbb^{1,d-1}$ can be distinguished as either future or past pointing, based on the sign of their time ($y^0$) component. In $\Rbb^{1,1}$ we have an additional refinement of space-like vectors into left- and right-moving, based on the sign of their space ($y^1$) component. Hence, any spacetime vector in $\Rbb^{1,1}$ falls into one of these four categories: future, past, left, right.

Our choice of metric is $(+,-)$, such that time-like events are always positively separated, in the sense that $(x-y)^2>0$ means that $x$ and $y$ are time-like. We will be working in dual momentum space such that a momentum vector is written as the difference of two points in dual momentum space: $p^\mu=x_{b}^\mu-x_a^\mu$. The momentum vectors are invariant under translations in dual momentum space, which allows us to fix the origin arbitrarily. Throughout the paper, We will use $x$ to denote the dual momenta corresponding to external particles, and $y$ for loop variables. Furthermore, we will find it useful to rotate into lightcone coordinates, such that we represent a point $y^\mu$ in dual momentum space by two real numbers $\lambda,\tilde\lambda$, which we define as
\begin{align}
    \lambda = y^0-y^1,\quad\tilde\lambda=y^0+y^1.
\end{align}
Although suggestive of spinor-helicity variables, 
$\lambda$ and $\tilde\lambda$ here are simply real coordinates parameterizing $\Rbb^{1,1}$ and do not carry spinor indices. In further analogy with spinor-helicity, we define
\begin{align}
    \<ij\> = \lambda_i-\lambda_j,\quad[ij]=\tilde\lambda_i-\tilde\lambda_j.
\end{align}
Using this notation we can see the factorization of squared distances in $\Rbb^{1,1}$:
\begin{align}
    (y_i-y_j)^2=\<ij\>[ij].
\end{align}
We will use the notation $y_i=(\lambda_i,\tilde\lambda_i)$ if the point $y_i^\mu\in\Rbb^{1,1}$ has lightcone coordinates $\lambda=\lambda_i, \; \tilde\lambda=\tilde\lambda_i$.

%%%%%%%%%%%%%%%%%%%%%%%%%%%%%%%%%%%%%%%%
\section{A Lightcone Geometry in Two Dimensions}\label{sec:def}

We start by defining a tree-level geometry $\Delta^{(0)}(x_a,x_b)$ as the space of time-like future-moving momenta. For notational simplicity, we will denote $p=x_b-x_a$, and reserve the numeric subscript for loop variables. Using lightcone coordinates, future moving momenta satisfy $\<ab\><0, [ab]<0$. If we use translation invariance to fix $x_a$ (say at the origin), then this geometry has a canonical form
\begin{align}
    \Omega(\Delta^{(0)}(x_a,x_b))=\dd\log\<ab\>\wedge\dd\log[ab]=\frac{\dd\lambda_b\wedge\dd\tilde\lambda_b}{\<ab\>[ab]}=\frac{\dd^2p}{p^2}.
\end{align}
Thus, the canonical form of this geometry recovers the tree-level two-point function $1/p^2$.
\begin{figure}
    \centering
    \includegraphics[width=0.5\linewidth]{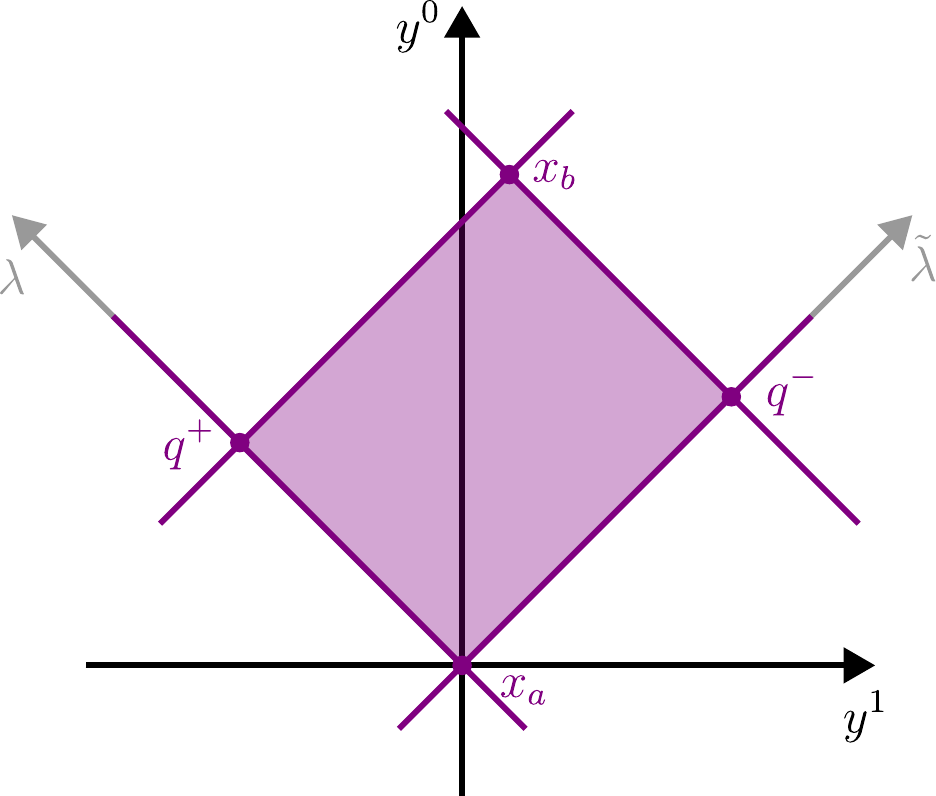}
    \caption{The one-loop geometry $\Delta(x_a,x_b)$.}
    \label{fig:1-loop}
\end{figure}

Next, we will define the lightcone geometries which capture the loop-level structure. Typically, in the lightcone geometry framework, one would define the $L$-loop object as a fibration over the tree-level geometry (see appendix \ref{sec:appendix}). Here, however, we fix some point $(x_a,x_b)$ from $\Delta^{(0)}$ (hence it satisfies $(x_a-x_b)^2>0$), and treat it as external kinematic input. This is equivalent to factoring out the tree-level geometry and considering only the loop-level factor, which makes the dual conformal invariance (DCI) more transparent. This is analogous to $\Ncal=4$ SYM, where it is natural to consider the $L$-loop amplitude divided by a tree-level MHV amplitude.

We now consider the region $y\in\Rbb^{1,1}$ which is time-like separated from both $x_a$ and $x_b$. This region naturally splits up into a compact part $\Delta(x_a,x_b)$ `between' $x_a$ and $x_b$, and a non-compact part. This can be seen in figure \ref{fig:1-loop}. For simplicity, we will sometimes denote $\Delta\equiv \Delta(x_a,x_b)$. The region $\Delta$ is precisely the \emph{causal diamond} generated by $x_a$ and $x_b$. We see that two new vertices arise at the intersection of both lightcones. Explicitly, we have
\begin{align}
    q^+=\frac{1}{2}\begin{pmatrix} x_a^0+x_a^1+x_b^0-x_b^1\\x_a^0+x_a^1-x_b^0+x_b^1 \end{pmatrix},\quad q^-=\frac{1}{2}\begin{pmatrix}
        x_b^0+x_b^1+x_a^0-x_a^1\\x_b^0+x_b^1-x_a^0+x_a^1
    \end{pmatrix}.
\end{align}
Or, in terms of lightcone coordinates,
\begin{align}
    &q^+=(\lambda_b, \tilde\lambda_a),\quad q^-=(\lambda_a , \tilde\lambda_b).
\end{align}
We can define the compact region $\Delta$ in terms of lightcone coordinates as
\begin{align}\label{eq:Delta-def-lambda}
    \Delta(x_a,x_b)=\{y=(\lambda, \tilde\lambda)\in\Rbb^{1,1} \ | \ \lambda_a\leq\lambda\leq\lambda_b,\; \tilde\lambda_a\leq\tilde\lambda\leq\tilde\lambda_b\}.
\end{align}
From this we can clearly see that $\Delta$ is a rectangle: it is the product of a line segment $\lambda_a\leq\lambda\leq\lambda_b$, and a line segment $\tilde\lambda_a\leq\tilde\lambda\leq\tilde\lambda_b$. Using this observation, we immediately find the canonical form to be
\begin{align}
    \Omega(\Delta(x_a,x_b))=\dd\log\frac{\lambda-\lambda_a}{\lambda-\lambda_b}\wedge\dd\log\frac{\tilde\lambda-\tilde\lambda_a}{\tilde\lambda-\tilde\lambda_b} = \frac{(x_a-x_b)^2 \dd^2y}{(y-x_a)^2(y-x_b)^2}.
\end{align}

\begin{figure}
    \centering
    \includegraphics[width=0.5\linewidth]{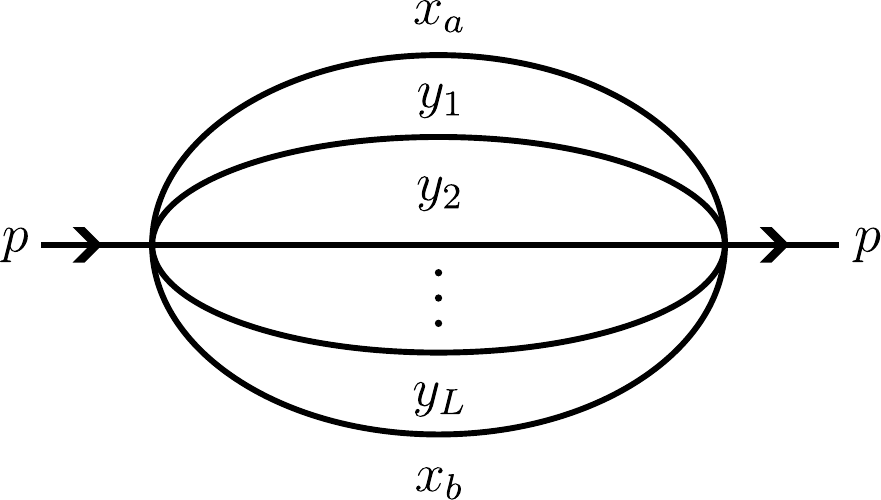}
    \caption{The $L$-loop banana graph.}
    \label{fig:banana}
\end{figure}

To generalize to higher loops, we define the $L$-loop lightcone geometry as
\begin{align}
    \Delta^{(L)}(x_a,x_b)\coloneqq \{(y_1,y_2,\ldots,y_L)\in[\Delta(x_a,x_b)]^L \ | \ (y_i-y_j)^2\geq 0\}.
\end{align}
This geometry is the space of $L$ mutually time-like separated points inside $\Delta$. A list of $L$ causally connected events is naturally ordered past to future, based on their $y^0$ component. This suggests a triangulation of $\Delta^{(L)}$ as 
\begin{align}
    \Delta^{(L)}=\bigcup_{\sigma\in S_L}\Delta_{\sigma(1)\cdots\sigma(L)}\equiv \bigcup_{\sigma \in S_L} \Delta_\sigma,
\end{align}
where we define
\begin{align}
    \Delta_{i_1,\ldots,i_L}\coloneqq \{(y_1,\ldots, y_L)\in\Delta^L \ | \ &\lambda_a\leq\lambda_{i_1}\leq\lambda_{i_2}\leq\cdots\leq\lambda_{i_L}\leq\lambda_b\notag\\
    &\tilde\lambda_a\leq\tilde\lambda_{i_1}\leq\tilde\lambda_{i_2}\leq\cdots\leq\tilde\lambda_{i_L}\leq\tilde\lambda_b\}.
\end{align}
Similar to the one-loop case, the $\lambda$ and $\tilde\lambda$ directions decouple, and $\Delta_{\sigma}$ is geometrically equivalent to the product of two $L$-simplices. The canonical form is given by
\begin{align}
    \Omega(\Delta_\sigma) &= \frac{\<ab\>\dd\lambda_{\sigma(1)}\wedge\cdots \wedge\dd\lambda_{\sigma(L)}}{\<a \sigma(1)\>\<\sigma(1)\sigma(2)\>\cdots\<\sigma(L)b\>} \wedge \frac{[ab]\dd\tilde\lambda_{\sigma(1)}\wedge\cdots \wedge\dd\tilde\lambda_{\sigma(L)}}{[a \sigma(1)][\sigma(1)\sigma(2)]\cdots[\sigma(L)b]} \\
    &=\frac{(x_a-x_b)^2\dd^2y_1\wedge\cdots\wedge\dd^2y_L}{(x_a-y_{\sigma(1)})^2(y_{\sigma(1)}-y_{\sigma(2)})^2\cdots (y_{\sigma(L)}-x_b)^2}.
\end{align}
We recognize this integrand as the massless $L$-loop banana graph, which is depicted in figure \ref{fig:banana}, with some additional factor $(x_a-x_b)^2$ in the numerator. The addition of this numerator makes the integrands \emph{dual conformal invariant}. This can be seen from the invariance under the inversion operator $\Ical(x^\mu)=x^\mu/x^2$.

\paragraph{Alternative Formulae.} We take this moment to briefly highlight a few equivalent ways in which the canonical form of the lightcone geometries can be written, which we will call upon in later sections. First, we note that we can write the one-loop canonical form as
\begin{align}\label{eq:1loop-dlog}
    \Omega(\Delta)=\pm\dd\log\frac{(y-x_a)^2}{(y-q^\pm)^2}\wedge\dd\log\frac{(y-x_b)^2}{(y-q^\pm)^2}.
\end{align}
The overall sign in front is equivalent to the sign in the superscript of $q$. This is reminiscent of the planar 4-point integrand for $\Ncal=4$ SYM, which can be written as \cite{Ferro:2023qdp}
\begin{align}\label{eq:N=4-1loop}
    \Omega(\Delta_4^{\Ncal=4\text{ SYM}})=\pm\dd\log\frac{(y-x_1)^2}{(y-q^\pm)^2}\wedge\dd\log\frac{(y-x_2)^2}{(y-q^\pm)^2}\wedge\dd\log\frac{(y-x_3)^2}{(y-q^\pm)^2}\wedge\dd\log\frac{(y-x_4)^2}{(y-q^\pm)^2}.
\end{align}
Next, we will find it valuable to consider various factorization of the $L$-loop canonical form into lower-loop functions. These formulae are inspired by the fibrational method which we review in appendix \ref{sec:appendix}. We can interpret $\Delta_{12\cdots L}$ by fibrating $y_L$ over $\Delta_{12\cdots L-1}$. It is clear that $y_L$ is allowed to be in the region $\Delta(y_{L-1},x_b)$, and hence
\begin{align}
    \Omega(\Delta_{12\ldots L}(x_a,x_b))=\Omega(\Delta_{12\ldots L-1}(x_a,x_b))\wedge\Omega(\Delta_{y_L}(y_{L-1},x_b)).
\end{align}
Here we use the notation $\Omega(\Delta_{y_L}(y_{L-1},x_b))$ to emphasize that it is the canonical form of a one-loop lightcone geometry in the variable $y_L$. Iterating this process leads to a `fibrations of fibrations' formula given by
\begin{align}
    \Omega(\Delta_{12\ldots L}(x_a,x_b))=\Omega(\Delta_{y_1}(x_a,x_b))\wedge\Omega(\Delta_{y_2}(y_1,x_b))\wedge\cdots\wedge \Omega(\Delta_{y_L}(y_{L-1},x_b)).
\end{align}
From a different perspective, we can equivalently fibrate $L-1$ loop variables over $\Delta(x_a,x_b)$. This leads to a similar formula of the form
\begin{align}
    \Omega(\Delta_{12\ldots L}(x_a,x_b))=\Omega(\Delta_{12\ldots L-1}(x_a,y_L))\wedge\Omega(\Delta_{y_L}(x_a,x_b)).
\end{align}
The iterative structure of these fibrational product formulae will prove to be useful when integrating the integrands in section \ref{sec:integrating}.

\subsection{Internal Boundaries}\label{sec:internal-boundaries}

We have seen that the full lightcone geometry $\Delta^{(L)}$ can be triangulated into $L!$ ``loop ordered'' regions $\Delta_\sigma$. 
Each of these $\Delta_\sigma$ is a convex polytope which can be decomposed as the Cartesian product of two $L$-simplices (one in $\lambda$-variables, and one in $\tilde\lambda$-variables). We note that this triangulation has no spurious boundaries that cancel, and that the interiors of two distinct loop ordered regions do not overlap. Two regions $\Delta_\sigma$ and $\Delta_{\sigma'}$ share at most a codimension-2 boundary, and this happens exactly when $\sigma$ and $\sigma'$ differ by an adjacent transposition. For example, $\Delta_{123\ldots L}$ and $\Delta_{213\ldots L}$ share a codimension-2 boundary as $\<12\>=[12]=0$.

There is a special two-dimensional boundary which is shared by all loop ordered regions, which is geometrically a one-loop geometry $\Delta(x_a,x_b)$. This happens when all $\lambda_i=\lambda_j, \tilde\lambda_i=\tilde\lambda_j \ \forall \ i,j$. Let us focus on a vertex of this two-dimensional boundary, say the point $\lambda_1=\ldots=\lambda_L=\lambda_a,\;\tilde\lambda_1=\ldots=\tilde\lambda_L=\tilde\lambda_a$, i.e. we localize all loop variables on $x_a$. The value of the residue at this vertex depends on the order in which we take the residues. For instance, consider the sequence of residues (note that the order of residues should be read right to left)
\begin{align}\label{eq:internal-boundary}
    \Res_{y_L=x_a}\;\Res_{y_{L-1}=y_{L}}\cdots\Res_{y_2=y_3}\;\Res_{y_1=y_2} \Omega(\Delta^{(L)}),
\end{align}
where we use the shorthand notation $\Res_{y_i=y_j}$ for the double residue $\smash{\Res_{\lambda_i=\lambda_j}\;\Res_{\tilde\lambda_i=\tilde\lambda_j}}$.
There are $2^{L-1}$ permutations which give a nonzero contribution to this sequence of residues\footnote{The counting is trivial. We can iteratively place points on a line, where each new point has to be either to the right or to the left of all previous points. Since each new point (other than the first one) has two options, there are a total of $2^{L-1}$ permutations which can be constructed in this way. These are precisely the permutations which contribute to this residue.}, and hence the result of \eqref{eq:internal-boundary} will be $\pm2^{L-1}$. However, a different sequence of residues can give us a different value, even though we still localize on the same vertex. For example,
\begin{align}
    \Res_{y_L=x_a}\cdots\Res_{y_2=x_a}\;\Res_{y_1=x_a} \Omega(\Delta^{(L)})=1,
\end{align}
since only $\Delta_{12\ldots L}$ will contribute.

This contradicts the usual definition of positive geometries, which only allows a residue of $\pm1$ at vertices. This is a known quirk of higher-loop amplituhedra, and it goes by the name of \emph{internal boundaries} \cite{Dian:2022tpf}. To allow for the presence of these internal boundaries, we need to adopt a slightly nonstandard definition of `positive geometries', which we review in appendix \ref{sec:appendix}.

%%%%%%%%%%%%%%%%%%%%%%%%%%%%%%%%%%%%%%%%
\section{Boundary of the Amplituhedron}\label{sec:boundary}
\begin{figure}
    \centering
    \includegraphics[width=0.5\linewidth]{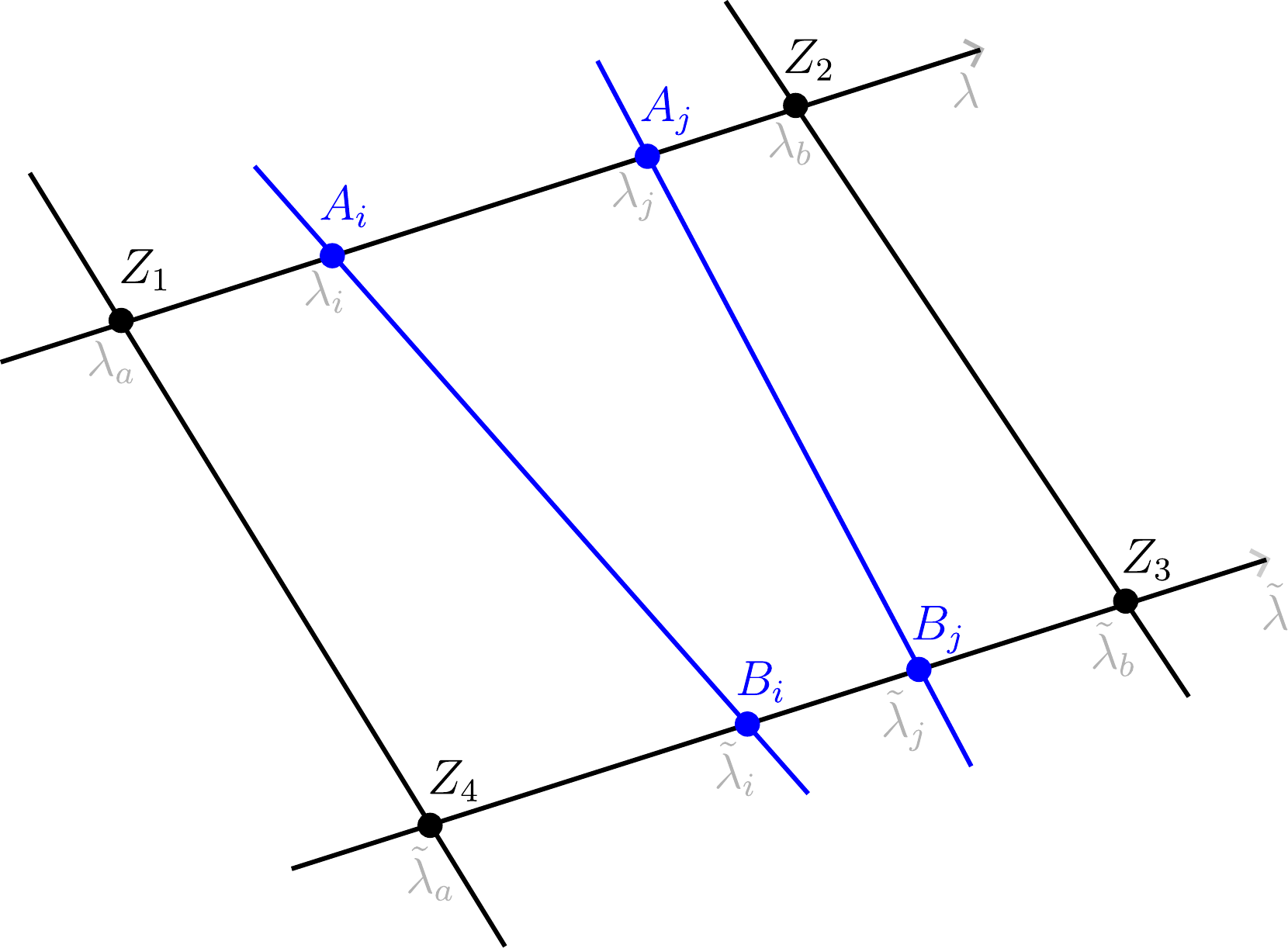}
    \caption{The configuration of points and lines in momentum twistors space which specify the boundary of the Amplituhedron given by $\<A_iB_i12\> = \<A_iB_i34\>=0$. }
    \label{fig:amp-boundary}
\end{figure}

In this section we will show how the lightcone geometry $\Delta^{(L)}$ connects to the amplituhedron. The $L$-loop amplituhedron $\Acal_4^{(L)}$ for planar 4-point MHV integrands in $\Ncal=4$ SYM is defined as follows. We fix four momentum twistors $Z_1, Z_2 ,Z_3, Z_4 \in \Pbb^3$ such that $\smash{\det\mat{Z_1 & Z_2 & Z_3 & Z_4}}\equiv\<1234\> >0$. We then define $\Acal_4^{(L)}$ to be the $4L$-dimensional space of $L$ lines $(A_i\;B_i)$, $i=1,\ldots, L$, which satisfy
\begin{align}
    &\<A_iB_i12\> >0, \quad \<A_iB_i23\>>0,\quad \<A_iB_i34\> >0,\quad \<A_iB_i14\>>0\\
    &\<A_iB_i13\><0,\quad \<A_iB_i24\><0,\quad \forall \ i=1,\ldots,L,
\end{align}
as well as the mutual positivity condition
\begin{align}
    \<A_i B_iA_jB_j\>>0, \quad \forall \ i,j=1,\ldots,L.
\end{align}
We now consider a specific boundary of this amplituhedron where we restrict $\<A_iB_i12\> = \<A_iB_i34\>=0$. These constraints can be solved by taking $A_i$ on the line $(Z_1 \;Z_2)$, and $B_i$ on the line $(Z_3 \; Z_4)$. The inequalities $\<A_iB_i14\> >0$ then restricts $A_i$ to be in between $Z_1$ and $Z_2$ on this line. Similarly, the inequality $\<A_iB_i23\>>0$ forces $B_i$ to be in between $Z_3$ and $Z_4$. The mutual positivity condition $\<(AB)_i(AB)_j\>>0$ now has two solutions for the ordering of the points: $(Z_1< A_i<A_j <Z_2)\times(Z_4< B_i< B_j< Z_3)$, or $(Z_1< A_j<A_i <Z_2)\times(Z_4< B_j< B_i< Z_3)$, where we use the notation $X<Y<Z$ to denote the ordering of points on a line. These two orders are analogous to the triangulation of $\Delta^{(L)}$ into loop-ordered regions. We depict the first case in figure \ref{fig:amp-boundary}.

If we suggestively parametrize the line $(Z_1\;Z_2)$ by a parameter $\lambda$ and $(Z_3\;Z_4)$ by a parameter $\tilde\lambda$, then the relation between this boundary of the $L$-loop amplituhedron to $\Delta^{(L)}$ becomes obvious. To make this interpretation precise, we choose our parameters $\lambda$ and $\tilde\lambda$ such that $Z_1$ corresponds to $\lambda=\lambda_a$, $Z_2 \to \lambda_b,\;Z_4\to\tilde\lambda_a,\;Z_3\to\tilde\lambda_b$, in which case the inequalities which describe this boundary of $\Acal_4^{(L)}$ are equivalent to those for $\Delta^{(L)}$. With this identification of $\Delta^{(L)}$ as a boundary of $\Acal_4^{(L)}$, the internal boundaries which we discussed in section \ref{sec:def} are then equivalent to the internal boundaries of the amplituhedron which were observed in \cite{Dian:2022tpf}.

\begin{figure}
    \centering
    \includegraphics[width=0.5\linewidth]{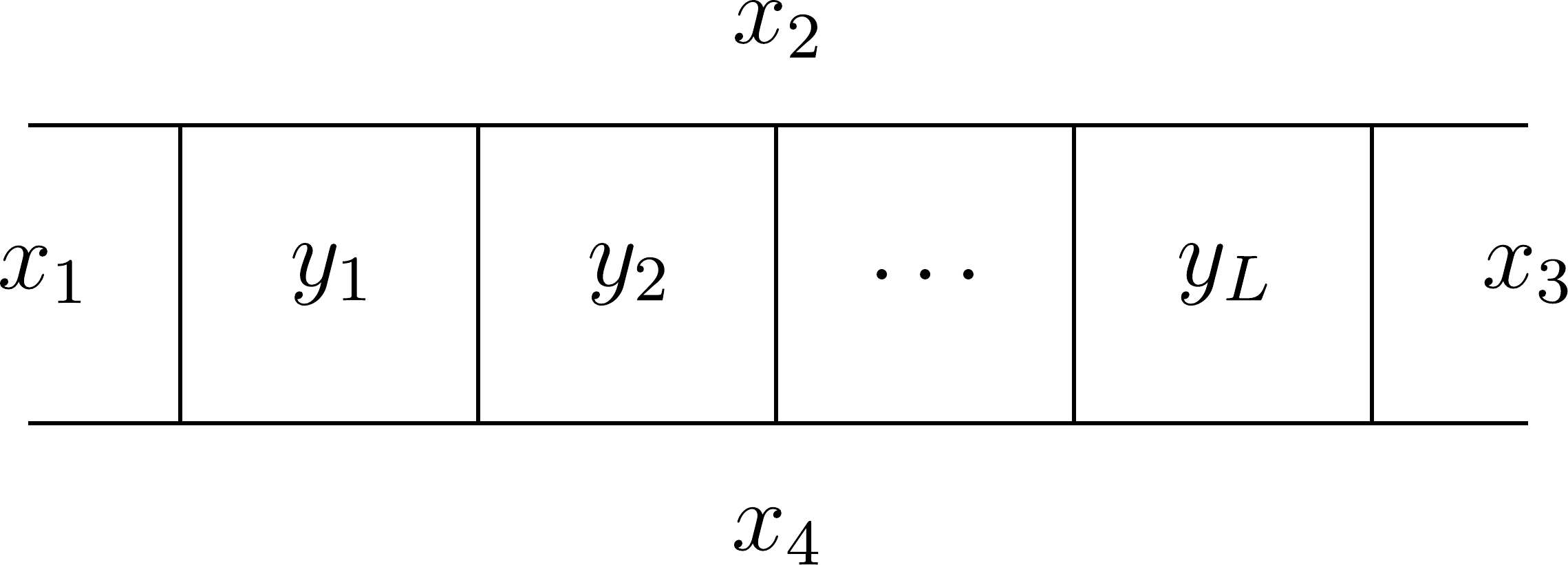}
    \caption{The $L$-loop ladder graph.}
    \label{fig:ladder}
\end{figure}

There is a standard way to translate lines in momentum twistor space into points in dual momentum space: $(Z_1 \;Z_4)\to x_1,\;(Z_1\; Z_2)\to x_2,\;(Z_2\; Z_3)\to x_3,\;(Z_3\; Z_4)\to x_4 $, which form a null-polygon (i.e. all edges of the polygon are light-like). Similarly, the loop variables map to $(A_i\;B_i)\to y_i$, and the diagonals $(Z_1\;Z_3)\to q^+_{1234},\;(Z_2\;Z_4)\to q^-_{1234}$ map to quadruple intersection points of the lightcones of $x_1,x_2,x_3,$ and $x_4$. In this language, the 4-point MHV amplituhedron $\Acal_4^{(L)}$ is a lightcone geometry in $\Rbb^{2,2}$ \cite{Ferro:2023qdp}. In terms of dual momentum variables, the boundary we are considering is given by $(y_i-x_2)^2=(y_i-x_4)^2=0$. From equations \eqref{eq:1loop-dlog} and \eqref{eq:N=4-1loop} it is easy to see that this residue indeed reduces $\Omega(\Delta_4^{\Ncal=4\text{ SYM}})$ into $\Omega(\Delta^{(1)})$. In general, the only DCI Feynman integrals in four dimensions which survive this cut are the so-called `ladder graphs', which we depict in figure \ref{fig:ladder}.
On a functional level, the only propagators in the ladder diagram which survive the residue at $(y_i-x_2)^2=(y_i-x_4)^2=0$ are 
\begin{align}
    \frac{1}{(x_1-y_1)^2(y_1-y_2)^2\cdots (y_L-x_3)^2},
\end{align}
which reproduces the pole structure of $\Omega(\Delta_{12\ldots L})$. Hence, it also becomes clear from the dual space picture that this boundary of the amplituhedron reproduces the banana graphs of figure \ref{fig:banana}. The conceptual difference is that the momenta are now still interpreted as 4-vectors in $\Rbb^{2,2}$, which are constrained to lie on a two-dimensional surface defined by the intersection of the lightcones of $x_2$ and $x_4$.

%%%%%%%%%%%%%%%%%%%%%%%%%%%%%%%%%%%%%%%%
\section{Integrating the Integrand}\label{sec:integrating}
We have defined the dual conformal invariant $L$-loop integrands $\Omega(\Delta^{(L)})$. In this section we will investigate what the integrated answer looks like, which we denote
\begin{align}
    A^{(L)}=\int \Omega(\Delta^{(L)}).
\end{align}
The $L!$ loop-ordered regions $\Delta_\sigma$ all integrate to the same result, as they are related by a simple relabeling of the loop variables, and hence we can simply consider integrating $\Delta_{12\ldots L}$. Much like its counterpart for $\Ncal=4$ SYM, the dual conformal invariance is broken after integration. The only kinematic invariant is $X\equiv (x_a-x_b)^2$, and hence we cannot construct any DCI cross-ratios and there is no nontrivial remainder function. We will use dimensional regularization with $D=2-2\epsilon$.

Let us first look at the one-loop case: the massless scalar bubble in 2D. We state the following result valid in general dimensions \cite{Smirnov:2004ym}:
\begin{align}\label{eq:gen-bubble}
    J^D_{\lambda_1,\lambda_2}(q^2)\equiv\int \frac{\dd^D k}{(k^2)^{\lambda_1}((q-k)^2)^{\lambda_2}} = \frac{\pi^{D/2}}{(q^2)^{\lambda_1+\lambda_2-D/2}}\frac{\Gamma(\lambda_1+\lambda_2-D/2)\Gamma(D/2-\lambda_1)\Gamma(D/2-\lambda_2)}{\Gamma(\lambda_1)\Gamma(\lambda_2)\Gamma(D-\lambda_1-\lambda_2)}.
\end{align}
We are interested in the case where $D=2-2\epsilon$, $\lambda_1=\lambda_2=1$, $q^2=(x_a-x_b)^2$. We find
\begin{align}
    A^{(1)}(x_a,x_b)&=\int\frac{(x_a-x_b)^2\dd^{2-2\epsilon} y}{(y-x_a)^2(y-x_b)^2}=(x_a-x_b)^2 J^{2-2\epsilon}_{1,1}((x_a-x_b)^2) \\&= \pi^{1-\epsilon} ((x_a-x_b)^2)^{-\epsilon}\frac{\Gamma(-\epsilon)^2\Gamma(1+\epsilon)}{\Gamma(-2\epsilon)}\\
    &=c_1(\epsilon) X^{-\epsilon},
\end{align}
where we use $X=(x_a-x_b)^2$, and
\begin{align}
    c_1(\epsilon) = \pi^{1-\epsilon}\frac{\Gamma(-\epsilon)^2\Gamma(1+\epsilon)}{\Gamma(-2\epsilon)}.
\end{align}

We can use the fibrational formulae from section \ref{sec:def} to find a recursive structure for higher loop results. Let us start with the two-loop function, which is the massless sunset. We know that
\begin{align}
    \Omega(\Delta_{12}(x_a,x_b))=\Omega(\Delta_{y_2}(x_a,x_b))\wedge\Omega(\Delta_{y_1}(x_a,y_2)).
\end{align}
Hence, we find that
\begin{align}
    A^{(2)}&=\int\Omega(\Delta_{12}(x_a,x_b))=\int \dd^{2-2\epsilon} y_2 \Omega(\Delta_{y_2}(x_a,x_b))\int \dd^{2-2\epsilon}y_1 \Omega(\Delta_{y_1}(x_a,y_2))\\
    &=\int \dd^{2-2\epsilon} y_2 \Omega(\Delta_{y_2}(x_a,x_b)) A^{(1)}(x_a,y_2)\\
    &=c_1(\epsilon) (x_a-x_b)^2\int \frac{1}{[(y-x_a)^2]^\epsilon}\frac{\dd^{2-2\epsilon} y}{(y-x_a)^2(y-x_b)^2}\\ &=c_1(\epsilon) X J^{2-2\epsilon}_{1+\epsilon,1}(X).
\end{align}
We find explicitly that
\begin{align}
    A^{(2)}(X)=c_2(\epsilon) X^{-2\epsilon},
\end{align}
where
\begin{align}
    c_2(\epsilon)=c_1(\epsilon) \pi^{1-\epsilon}\frac{\Gamma(-\epsilon)\Gamma(-2\epsilon)\Gamma(1+2\epsilon)}{\Gamma(-3\epsilon)\Gamma(1+\epsilon)}=\pi^{2-2\epsilon}\frac{\Gamma(-\epsilon)^3\Gamma(1+2\epsilon)}{\Gamma(-3\epsilon)}.
\end{align}
This fibrational approach generalizes to higher loops as 
\begin{align}
    A^{(L)}=X\int  \frac{A^{(L-1)}(x_a,y)}{(y-x_a)^2(y-x_b)^2}\dd^{2-2\epsilon} y.
\end{align}
This iterated integral structure implies that if
\begin{align}
    A^{(L-1)}(X)=c_{L-1}(\epsilon) X^{-(L-1) \epsilon},
\end{align}
then
\begin{align}
    A^{(L)}(X)= X c_{L-1}(\epsilon) J^{2-2\epsilon}_{1+(L-1)\epsilon,1}(X)&=c_{L-1}(\epsilon)\pi^{1-\epsilon}\frac{\Gamma(-\epsilon)\Gamma(-L\epsilon)\Gamma(1+L\epsilon)}{\Gamma(-(L+1)\epsilon)\Gamma(1+(L-1)\epsilon)} X^{-L\epsilon} \\&\equiv c_L(\epsilon) X^{-L\epsilon}.
\end{align}
We thus have a recursion relation for $c_L$ given by
\begin{align}
    c_0(\epsilon)=1,\quad c_L(\epsilon) = c_{L-1}(\epsilon)\pi^{1-\epsilon}\frac{\Gamma(-\epsilon)\Gamma(-L\epsilon)\Gamma(1+L\epsilon)}{\Gamma(-(L+1)\epsilon)\Gamma(1+(L-1)\epsilon)},
\end{align}
which can easily be solved to give
\begin{align}
    c_L(\epsilon)=\pi^{L(1-\epsilon)} \frac{\Gamma(-\epsilon)^{L+1}\Gamma(1+L\epsilon)}{\Gamma(-(L+1)\epsilon)}.
\end{align}
This gives the precise result for the massless $L$-loop banana graph which is exact in $\epsilon$. 

\subsection{Divergences}
Expanding the one-loop function in $\epsilon$ yields
\begin{align}
    A^{(1)}(x_a,x_b)&=\int \frac{(x_a-x_b)^2 \dd^{2-2\epsilon}y}{(y-x_a)^2(y-x_b)^2}=\frac{\Gamma(-\epsilon)^2\Gamma(1+\epsilon)}{\Gamma(-2\epsilon)}\pi^{1-\epsilon}X^{-\epsilon}\\&=-\frac{2\pi}{\epsilon} +2\pi\left[\log X +\gamma_E+\log\pi\right]+ \Ocal(\epsilon),
\end{align}
The $1/\epsilon$ pole is an \emph{IR divergence}. This divergence corresponds to a soft limit where $y$ approaches one of the two vertices $x_a$ or $x_b$ of $\Delta$. 

We recall that the $L$-loop function factorizes as
\begin{align*}
    A^{(L)}=c_L(\epsilon) X^{-L\epsilon}.
\end{align*}
The function $c_L$ is independent of the kinematic variable $X$ and contains the full IR divergence. The leading divergence is of the form
\begin{align}
    A^{(L)}=\frac{(-1)^L(L+1)\pi^L}{\epsilon^L}+\Ocal\left(\frac{1}{\epsilon^{L-1}}\right).
\end{align}
Alternatively, we note that we can write the $L$-loop result in terms of the $L$\textsuperscript{th} power of the 1-loop result as
\begin{align}
    A^{(L)} = \frac{(L+1)}{2^L} R_L(\epsilon) (A^{(1)})^L,
\end{align}
where
\begin{align}
    R_L(\epsilon)=\frac{\Gamma(1-\epsilon)\Gamma(1+L\epsilon)}{\Gamma(1-(L+1)\epsilon)}\left(\frac{\Gamma(1-2\epsilon)}{\Gamma(1-\epsilon)\Gamma(1+\epsilon)}\right)^L.
\end{align}
The function $R_L(\epsilon)$ satisfies $R_L(0)=1$, and thus only contributes to the subleading poles. We see that the full divergence structure and all kinematic dependence is captured by the $L$\textsuperscript{th} power of the one-loop function. This is analogous to \emph{IR exponentiation}, although resumming will not actually give rise to an exponential, as we will see in the next section. 

\subsection{Non-Perturbative Results}
We are interested in the quantity we get once we resum all the loop functions
\begin{align}
    A(X;\epsilon)=\sum_{L=0}^\infty g^LA^{(L)}=\sum_{L=0}^\infty g^L c_L(\epsilon) X^{-L\epsilon}.
\end{align}
We introduce the variable
\begin{align}
    z= g \pi^{1-\epsilon}\Gamma(-\epsilon) X^{-\epsilon},
\end{align}
such that
\begin{align}
    g^L A^{(L)}=  z^L \Gamma(-\epsilon)\frac{\Gamma(1+L\epsilon)}{\Gamma(-\epsilon-L\epsilon)}.
\end{align}
The infinite sum can be solved in terms of the \emph{Fox--Wright function} \cite{karp2019foxwrightfunctionnearsingularity}, which is a generalization of hypergeometric functions that naturally arises in sums with Gamma functions whose arguments grow linearly with $L$:
\begin{align}
    A(z;\epsilon) = \Gamma(-\epsilon)\times{}_2\Psi_1\left[\begin{array}{c} (1, \epsilon) \quad (1, 1) \\ (-\epsilon, -\epsilon) \\ \end{array}; z \right].
\end{align}
This sum converges for all finite $z$ when $\epsilon<0$.

Since $R_L(\epsilon)=1+\Ocal(\epsilon)$, it is natural to consider the approximation of $A^{(L)}$ when we set $R_L$ to 1. In this case we can resum to give a clean double pole:
\begin{align}
    \sum_{L=0}^\infty g^L \frac{(L+1)}{2^L} (A^{(1)})^L = \frac{1}{(1-\frac{g}{2}A^{(1)})^2}.
\end{align}
This provides a simple analytic expression which captures the leading IR behavior of the full non-perturbative result.

%%%%%%%%%%%%%%%%%%%%%%%%%%%%%%%%%%%%%%%%
\section{Large-\texorpdfstring{$L$}{L} Limit}\label{sec:large-L}

We will now consider what happens to the integrated $L$-loop banana graph in the limit when $L$ goes to infinity
\begin{align}
    \lim_{L\to\infty}\int \Omega(\Delta^{(L)})&=\lim_{L\to\infty} \int\frac{(x_a-x_b)^2\dd^2 y_1\cdots\dd^2 y_L}{(x_a-y_1)^2(y_1-y_2)^2\cdots (y_L-x_b)^2}.
\end{align}
We introduce a continuous curve $y(\sigma)$ such that $y_i=y(\sigma_i)$, $y(0)=x_a$, $y(1)=x_b$, where $\sigma_i=i\, \delta\sigma$ and $\delta\sigma=1/(L+1)$. When $L$ is large, the terms $(y_i-y_{i+1})^2$ become $(y(\sigma_i)-y(\sigma_i + \delta\sigma))^2\simeq\dot{y}^2(\sigma_i)\delta\sigma^2$, where we introduce $\dot{y}=\dd y/\dd\sigma$. We write the propagator factors in an exponential as
\begin{align}
    \frac{1}{(y_i-y_{i+1})^2}=e^{-\log (y_i-y_{i+1})^2}\simeq \frac{1}{\delta\sigma^2} e^{- \log \dot{y}^2(\sigma_i)}.
\end{align}
In the continuum limit, the product of propagators becomes an integral in the exponent
\begin{align}
    \prod_{i=0}^L\frac{1}{(y_i-y_{i+1})^2} \simeq \frac{1}{\delta\sigma^{2L}}e^{-\frac{1}{\delta\sigma}\int\dd \sigma \log \dot{y}^2(\sigma) }.
\end{align}
We want to integrate this exponential with the integration measure
\begin{align}
    \lim_{L\to\infty} \frac{\dd^2 y_1\cdots \dd^2 y_L}{\delta\sigma^{2L}},
\end{align}
which can be naturally interpreted as a discretized path integral measure $\Dcal^2 y$ for the scalar fields $y^0$ and $y^1$. We thus have the emergence of a path integral at large $L$:
\begin{align}
    \lim_{L\to\infty} \int \Omega(\Delta^{(L)}) \simeq\int\Dcal^2y(\sigma) e^{-\frac{1}{\delta\sigma}\int \dd\sigma\log \dot{y}^2}.
\end{align}
This resembles a worldline path integral with an effective Lagrangian given by the logarithm of the squared velocity. 

Since $1/\delta\sigma \sim L$, this integral is dominated by its saddle point in the large-$L$ limit. We have the action
\begin{align}
    S[y]=\int_0^1\dd\sigma \log\dot{y}^2,
\end{align}
which gives us the saddle point equations
\begin{align}
    \frac{\dd}{\dd\sigma} \left(\frac{\dot{y}}{\dot{y}^2}\right)=0.
\end{align}
This has the solution $\dot{y}=\text{const.}$ In this case $y(\sigma)$ is a straight line between $x_a$ and $x_b$, or, explicitly, $y=x_a + \sigma (x_b-x_a)$, which gives $\dot{y}=x_b-x_a$. We will call this the ``classical solution''. The classical action is then
\begin{align}
    S_{\text{class}}=\int_0^1\dd\sigma \log (x_a-x_b)^2= \log (x_a-x_b)^2. 
\end{align}
Thus, in the large-$L$ limit, the banana integrals behave as
\begin{align}
    \lim_{L\to\infty}\int\Omega(\Delta^{(L)}) \sim X^{-L}
\end{align}
at leading order.

\begin{figure}
    \centering
    \includegraphics[width=0.5\linewidth]{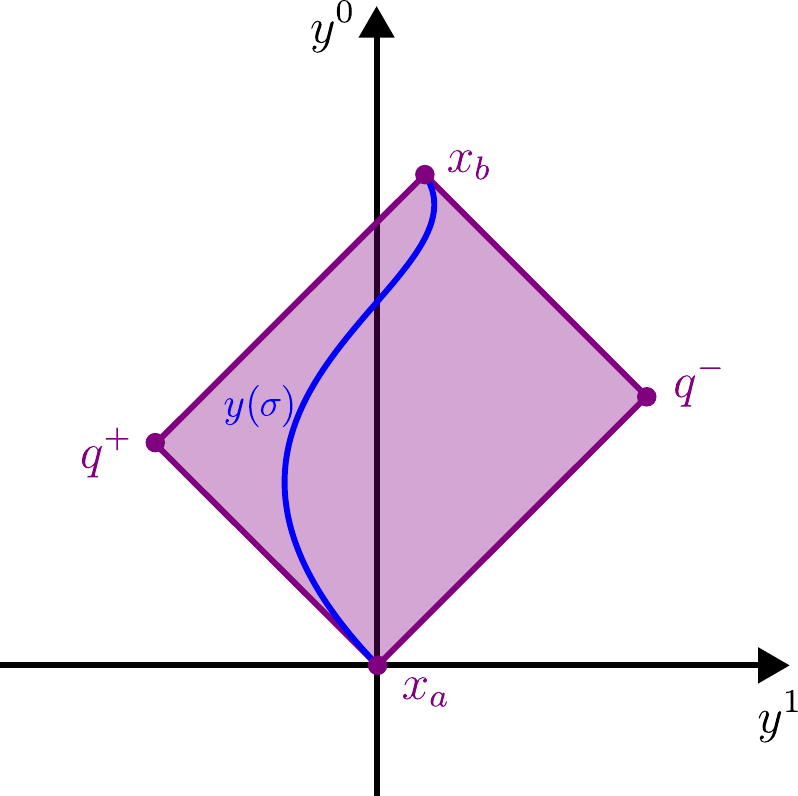}
    \caption{An example of a worldline $y(\sigma)$, which represents a point in $\Delta^{(\infty)}$. }
    \label{fig:worldline}
\end{figure}

Geometrically, in the $L\to\infty$ limit $\Delta^{(L)}$ approaches the infinite-dimensional space of all causal worldlines $y(\sigma)$ which start at $x_a$ and end at $x_b$. An example is given in figure \ref{fig:worldline}. From this perspective, it is not too surprising that the $L\to\infty$ limit reproduces a path integral\footnote{It should be noted that this set of curves specifies the geometry in this limit, but it does not specify the region of integration. This is analogous to the fact that the one-loop bubble is obtained by integrating $\Omega(\Delta)$ over all of $\Rbb^{1,1}$, not just over $\Delta$.}. 

The emergence of a path integral in the large-$L$ limit is reminiscent of the kind of dual theories which appear in holography at strong coupling. We take this as suggestive evidence that such a dual description may exist for the present geometry, though we leave a precise identification to future work.

%%%%%%%%%%%%%%%%%%%%%%%%%%%%%%%%%%%%%%%%
\section{Conclusions and Outlook}\label{sec:conclusions}

We have defined and studied a positive geometry $\Delta^{(L)}$, which arises both as a natural generalization of loop amplituhedra to two dimensions, and as a specific boundary of the $L$-loop 4-point amplituhedron for $\Ncal=4$ SYM. We interpret $\Delta^{(L)}$ within the framework of ``lightcone geometries''. 

We showed that the canonical form of $\Delta^{(L)}$ provides a family of dual conformal invariant integrands in two dimensions: the massless banana graphs. These functions can be integrated at all loop orders and resummed into a non-perturbative result. At large-$L$, we have seen the emergence of a path integral, giving strong evidence for the existence of a dual theory at strong coupling. 

A natural next step is the identification and study of this dual theory at strong coupling. This may be aided by the integrated $L$-loop function and its non-perturbative resummation. Furthermore, the emergence of a path integral at large loops is analogous to the continuum limit of fishnet theories \cite{Basso:2018agi}, and it would be interesting to see if this connection can be made precise.

It is particularly interesting to investigate this dual theory from a geometric point of view, especially in connection with the study of the \emph{dual amplituhedron}. The definition of a dual amplituhedron for $\Ncal=4$ SYM is one of the important open questions in the study of positive geometries, and should facilitate a connection to the dual weakly coupled string theory at strong 't Hooft coupling \cite{Arkani-Hamed:2014dca}. Since our loop ordered lightcone geometries are purely polytopal, it readily admits a description as a dual polytope. Our two-dimensional amplituhedron is therefore in a prime position to explicitly explore this correspondence between dual amplituhedra and dual physical theories. 

More broadly, $\Delta^{(L)}$ provides a natural and physically interesting toy model to study various aspects of the loop structure of the amplituhedron. As such, there are several avenues of investigation which might benefit from this simplification. As an example, the 4-point $L$-loop \emph{deformed amplituhedron} conjecturally describes loop integrands of $\Ncal=4$ SYM on the Coulomb branch \cite{Arkani-Hamed:2023epq}. This has a straightforward generalization to $\Rbb^{1,1}$, and presumably captures some massive version of the DCI banana graphs we studied in this paper. This geometry would additionally have an interpretation as a boundary of the four-dimensional deformed amplituhedron, suggesting that the relative simplicity of the two-dimensional setting may provide useful input for bootstrapping the canonical form for $\Ncal=4$ SYM. These investigations can be further aided by recent results on massive banana graphs at all loop orders \cite{Vanhove:2026fth}.

Additionally, it would be interesting to investigate \emph{negative geometries} \cite{Arkani-Hamed:2021iya} in $\Rbb^{1,1}$. By changing some of the positivity conditions on $(y_i-y_j)^2$ with negativity conditions, or no conditions at all, we find a set of positive geometries which describe the integrands for the logarithm $\log(A)$. The negative geometries which arise from our two-dimensional model are equivalent to certain boundaries of the negative geometries for $\Ncal=4$ SYM \cite{Arkani-Hamed:2021iya}, and are considerably simpler and more tractable. For instance, using the fibrational approach, one can find a simple formula for the canonical forms of any `trees of loops' configuration. It might be interesting to see if certain poorly understood properties of the negative geometries for $\Ncal=4$ SYM, such as the remarkable accuracy of the `trees of loops' approximation of the non-perturbative result, might become more transparent in this simpler setting.

Overall, the simplicity of this two-dimensional setting provides a controlled arena which allows us to probe the geometric and physical structure of loop amplituhedra. We hope to return to some of these directions in future work.

\section*{Acknowledgments}
I am grateful to Mattia Capuano, Gabriele Dian, Ross Glew, Livia Ferro, Karol Kampf, Tomasz \L{}ukowski, and David Podiv\'in for fruitful discussions. This work is supported by OP JAK \v{C}Z.02.01.01/00/22\_008/0004632.

\appendix

\section{Review of Positive Geometry}\label{sec:appendix}
In this appendix we will review some basic notions of positive geometries which are relevant to this paper. The term ``positive geometry'' describes a framework revolving around the study of certain geometric regions and their associated canonical form. Precise definitions of what constitutes a positive geometry and their canonical forms have been given in \cite{Arkani-Hamed:2017tmz} and \cite{Brown:2025jjg}. In this paper, our perspective is most closely aligned with the definition of \emph{weighted positive geometries} from \cite{Dian:2022tpf}, which generalizes the notion of positive geometries in \cite{Arkani-Hamed:2017tmz} to allow for the presence of \emph{internal boundaries}. For the purpose of the current paper it will, however, not be necessary to spend too much time on the mathematical intricacies of these definitions. We will continue with a physicist's level of rigor, and we explain the relevant notions at an appropriately intuitive, albeit slightly informal, level. We refer the reader to the aforementioned papers for a more in-depth discussion.

\paragraph{Definition.} We define a $d$-dimensional positive geometry $\Acal$ to be a closed, oriented, $d$-dimensional region inside the real slice of some ambient projective space. We require that $\Acal$ has boundaries of all codimension (that is, it has boundaries of dimension $d'$ for all $d'=0,1,\ldots,d-1$). The simplest examples of positive geometries include convex polytopes in $d$ dimensions. To the positive geometry we associate a canonical form $\Omega(\Acal)$, which is a $d$-form on the ambient projective space and which satisfies the following defining properties:
\begin{itemize}
    \item $\Omega(\Acal)$ has logarithmic singularities along the boundary surfaces of $\Acal$, and nowhere else.
    \item The residue of $\Omega(\Acal)$ at such a singularity yields the canonical form of the corresponding boundary. In particular, this means that all boundaries of a positive geometry are themselves a positive geometry. 
    \item A zero-dimensional positive geometry is a point, and its canonical form is $\Omega(\bullet)=\pm 1$, with the sign depending on orientation.
\end{itemize}
Typically, when considering real polytopes, the ambient projective space is left implicit. There is a natural embedding of $\Rbb^d$ into the real slice of $\Pbb^d$, which is understood in those cases.

\paragraph{Canonical Forms.} There are several ways in which the canonical form of a positive geometry can be found. We will highlight a few techniques which are important for the context of this paper. First, if we \emph{triangulate} the positive geometry as $\Acal=\Acal_1 \cup \Acal_2$, where the interiors of $\Acal_1$ and $\Acal_2$ do not intersect, then the canonical form of $\Acal$ can be found as 
\begin{align}
    \Acal=\Acal_1 \cup \Acal_2\quad\implies\quad\Omega(\Acal) = \Omega(\Acal_1)+\Omega(\Acal_2).
\end{align}
Secondly, when the positive geometry $\Acal$ is a product geometry $\Acal=\Bcal\times\Ccal$, where $\times$ denotes the Cartesian product, then the canonical form factorizes as
\begin{align}
    \Acal=\Bcal\times\Ccal \quad \implies \quad \Omega(\Acal)=\Omega(\Bcal)\wedge\Omega(\Ccal).
\end{align}
This factorization can be generalized beyond the Cartesian product via the \emph{fibrational method}. We consider a projection $\pi$ of $\Acal$ to a lower-dimensional positive geometry $\pi(\Acal)$. We can reconstruct the full geometry $\Acal$ by attaching a \emph{fiber} $f(x)\equiv\Acal\,\cap\,\pi^{-1}(x)$ to each point $x\in \pi(\Acal)$:
\begin{align}
    \Acal= \{(x,y) \ | \ x\in\pi(\Acal), \ y\in f(x)\}.
\end{align}
We can interpret the Cartesian product as a fibration with a constant fiber. We borrow the notation from the Cartesian product and write $\Acal=\pi(\Acal)\times f$, and it should be clear from context whether the fiber is constant or varies non-trivially as a function of $x\in\pi(\Acal)$. We say that the fiber is linear if its vertices are linear (more precisely, affine) functions of $x$, so that the polytope deforms without changing its combinatorial structure. As long as the fiber $f(x)$ is linear, then the canonical form factorizes as
\begin{align}\label{eq:fibration}
    \Acal=\pi(\Acal)\times f \quad \implies \quad \Omega(\Acal)=\Omega(\pi(\Acal))\wedge\Omega(f(x)).
\end{align}
In practice, when projecting a polytope, the fibers are often only piecewise linear, as the fiber can change shape discontinuously when its defining facets switch. In this case we triangulate $\pi(\Acal)$ into several \emph{chambers} such that the fibers are linear over each chamber. Mathematically, these chambers are examples of a \emph{polyhedral subdivision induced by the projection $\pi$}. We can then use a combination of the triangulation and fibrational methods to find the full canonical form\footnote{As an aside, we point out that it is known that the restriction to linear fibers $f$ is not necessary for the factorization to hold, and has been successfully used for various ``curvy'' loop amplituhedra for both $\Ncal=4$ SYM \cite{Ferro:2023qdp,Ferro:2024vwn} and ABJM \cite{He:2023rou,Lukowski:2023nnf}. However, it was recently shown in \cite{Ferro:2025pij} that there are certain higher-loop cases where the factorization of the canonical forms ceases to be valid. It is currently not known what the requirements are for the fibrational method to hold in the nonlinear case.}. These fibrational methods for polytopes have been used in \cite{Bartsch:2025mvy} to study the ABHY associahedron \cite{Arkani-Hamed:2017mur} and stringy generalizations thereof.

\begin{figure}
    \centering
    \includegraphics[width=0.4\linewidth]{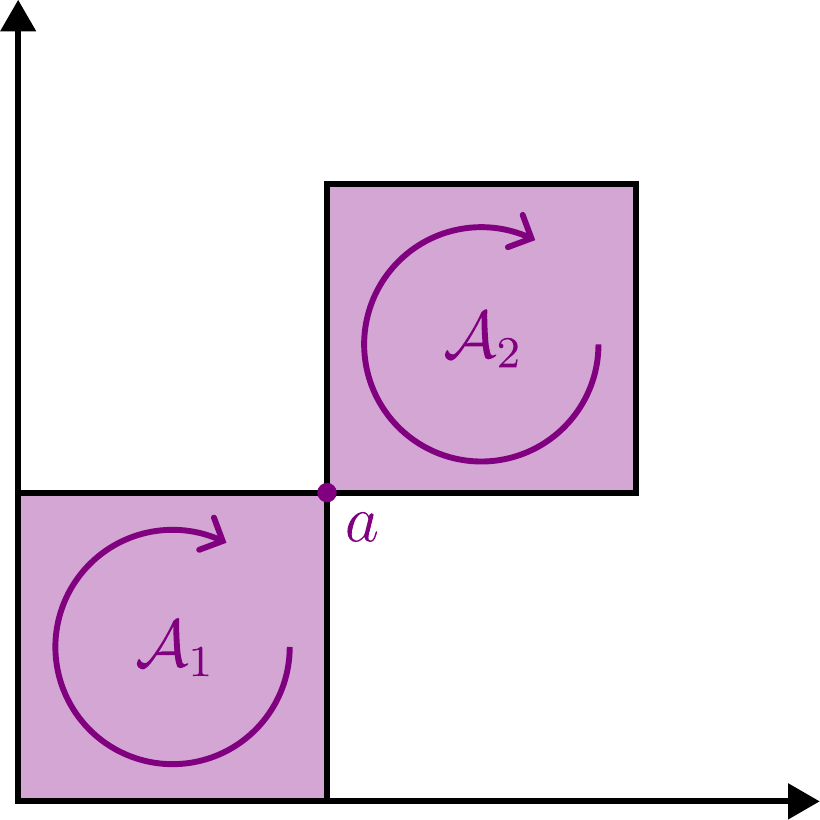}
    \caption{The geometry $\Acal=\Acal_1\cup\Acal_2$ has an internal boundary at the point $a$.}
    \label{fig:internal-boundary}
\end{figure}

\paragraph{Internal Boundaries.} At this point, our definition still coincides with the classical definition of positive geometries of \cite{Arkani-Hamed:2017tmz}. However, the geometries we consider in this paper \emph{do not} fit the above definition. This is due to the presence of \emph{internal boundaries}. These internal boundaries appear, for example, when two convex polytopes $\Acal_1$ and $\Acal_2$ share a boundary $\Bcal=\Acal_1\cap\Acal_2$ of codimension-2 (or more), but otherwise don't intersect. We can interpret the union $\Acal=\Acal_1\cup\Acal_2$ of these convex polytopes as a triangulation, and hence we expect its canonical form to just be the sum $\Omega(\Acal)=\Omega(\Acal_1)+\Omega(\Acal_2)$. In this case, when we consider the residue of the canonical form along this shared boundary, then we get a contribution from both $\Omega(\Acal_1)$ and $\Omega(\Acal_2)$. If $\Acal_1$ and $\Acal_2$ agree on their orientation, then they contribute to this boundary with the same sign, and hence we end up with \emph{twice} the canonical form of this boundary $\Omega(\Bcal)$. This is not consistent with the definition of the canonical form which we gave above, since the residue at a vertex of $\Bcal$ will now be $\pm 2$, rather than $\pm 1$. This is illustrated with a simple example in figure \ref{fig:internal-boundary}.

It is known that various geometries of interest have these kinds of internal boundaries. For instance, the loop amplituhedron for $\Ncal=4$ SYM is known to have them for $L\geq 2$. As such, they are not considered positive geometries in the literal sense following the definition, but they still fall under the same umbrella of ``positive geometry'' as a framework. The more appropriate mathematical definition is that of \emph{weighted positive geometries}, where boundary components are assigned multiplicities (or ``weights''), and the canonical form is defined such that its residues reproduce these weights \cite{Dian:2022tpf}. For the purpose of this paper, we simply relax the definition of positive geometries to allow for these internal boundaries. If a region can be triangulated by positive geometries following the definition above, then we give it a canonical form equal to the sum of the canonical forms of its constituents. We shall still refer to these geometries simply as ``positive geometries''. Most of the properties of positive geometries and their canonical forms we discussed are still valid, other than the fact that residues count the multiplicity. In this sense, internal boundaries behave as if they are counted multiple times.

\subsection{Lightcone Geometries}
We now briefly review a modern perspective on loop-level amplituhedra as \emph{lightcone geometries} in \emph{dual momentum space}, which has been developed for $\Ncal=4$ SYM and ABJM in \cite{Lukowski:2023nnf,Ferro:2023qdp,Ferro:2024vwn,Ferro:2025pij}. In particular, the geometry we study in this paper can be interpreted as a lightcone geometry in $\Rbb^{1,1}$, where many general features simplify drastically.

Given a point $x\in\Rbb^{d_1,d_2}$, we define the lightcone $\Ncal_x\coloneqq \{y\in\Rbb^{d_1,d_2} \ | \ (y-x)^2=0\}$. A \emph{lightcone geometry} in $\Rbb^{d_1,d_2}$ is a positive geometry whose facets (codimension-1 boundaries) are all lightcones. Typically, we start from a set of points $\{x_1,x_2,\ldots,x_n\}$ in $\Rbb^{d_1,d_2}$, and consider the space which satisfies $(y-x_i)^2\geq 0$ for all $x_i$. In many cases of physical interest, as for the geometry we consider in this paper, this region naturally decomposes into a compact part and a non-compact part. We denote the compact part as
\begin{align}
    \Delta(x_1,\ldots, x_n)\coloneqq \{y\in\Rbb^{d_1,d_2} \ | \ (y-x_i)^2\geq 0 \ \forall \ i=1,\ldots, n,\quad\text{compact}\}.
\end{align}
A concise characterization of this compact part is often rather intricate and relies on a certain \emph{sign-flip condition} on a sequence of invariants, which we will not review here. The facets of $\Delta(x_1,\ldots,x_n)$ are the lightcones $\Ncal_{x_1},\ldots,\Ncal_{x_n}$, and thus, assuming that $\Delta(x_1,\ldots,x_n)$ is a positive geometry, it falls in the category of lightcone geometries. Starting from $\Delta(x_1,\ldots,x_n)$, we define the \emph{$L$-loop lightcone geometry} as
\begin{align}
    \Delta^{(L)}(x_1,\ldots,x_n)\coloneqq \{(y_1,y_2,\ldots, y_L) \in [\Delta(x_1,\ldots,x_n)]^L \ | \ (y_i-y_j)^2 \geq0 \ \forall \ i,j=1,\ldots,L \}.
\end{align}
That is, it is the $L(d_1+d_2)$-dimensional space of $L$ mutually positively separated points in $\Delta(x_1,\ldots,x_n)$. The lightcone geometry $\Delta(x_1,\ldots, x_n)$ is sometimes referred to as the 1-loop lightcone geometry.

The relation between these lightcone geometries and loop amplituhedra can be understood using the fibrational method we reviewed in the previous section. The loop level (momentum) amplituhedron for both $\Ncal=4$ SYM \cite{Arkani-Hamed:2013jha,Damgaard:2019ztj, Ferro:2022abq, Ferro:2023qdp} and ABJM \cite{He:2021llb, Huang:2021jlh,He:2022cup,He:2023rou,Lukowski:2023nnf,Lukowski:2021fkf} can be interpreted as lightcone geometries in an equivalent way. We assume that we have a geometry $\Acal^{\text{tree}}$ which captures the color-ordered tree-level scattering of some theory. Each point in $\Acal^{\text{tree}}$ specifies the momenta of $n$ scattering particles in $\Rbb^{d_1,d_2}$, which can equivalently be cast into \emph{dual momentum variables}. These are variables $x_i$ such that the momenta are given by $p_i=x_{i+1}-x_i$. Each point in $\Acal$ thus specifies $n$ points $x_1,x_2,\ldots x_n$ in dual momentum space. We now use these points to define the lightcone geometry $\Delta^{(L)}(x_1,\ldots,x_n)$, and fibrate it over all points in $\Acal$. This defines the $L$-loop positive geometry
\begin{align}
    \Acal^{(L)} \coloneqq \Acal^{\text{tree}} \times\Delta^{(L)}(x_1,\ldots,x_n),
\end{align}
whose canonical form captures the planar $L$-loop integrand for the relevant theory. The compactness of the lightcone geometry reflects the dual conformal invariance of these integrands. The facets of these geometries are lightcones of the form $(y_i-x_j)^2=0$, corresponding to a forward limit, or $(y_i-y_j)^2=0$. This precisely matches the codimension-1 poles we expect from these integrands (factorizations manifest as boundaries of $\Acal^{\text{tree}}$).

Although the $L$-loop lightcone geometries are typically not polytopal, the factorization of canonical forms often still works in practice, after summing over the appropriate chambers $\mathfrak{c}_i^{\text{tree}}$ which triangulate $\Acal^{\text{tree}}$:
\begin{align}
    \Omega(\Acal^{(L)})=\sum_i\Omega(\mathfrak{c}_i^{\text{tree}})\wedge\Omega(\Delta_i^{(L)}).
\end{align}
In two spacetime dimensions, however, a lightcone $\Ncal_x$ factorizes into two lightrays. This renders the lightcone geometries linear, which makes the factorization transparent. Furthermore, the geometry which we consider in this paper only has a single chamber, and as such we can factor out a global tree-level contribution and study $\Delta^{(L)}$ as an independent quantity.

%%%%%%%%%%%%%%%%%%%%%%%%%%%%%%%%%%%%%%%%%%%
\bibliographystyle{nb}

\bibliography{bibliography}
	
\end{document}